\documentclass[useAMS,usenatbib]{mnras}
\usepackage{graphics}
\usepackage{graphicx}
\usepackage{amssymb}
\usepackage{amsfonts}
\usepackage{wasysym}
\usepackage{epsfig}
\usepackage{longtable}
\usepackage{times}
\usepackage{mathptmx}
\usepackage[usenames,dvipsnames]{color}
\usepackage[normalem]{ulem}
\usepackage{bm}

\title[Exploring the circumstellar environment of HR Car]
{
Exploring the multifaceted circumstellar environment of the Luminous Blue Variable  HR~Carinae
}
\author[C. Buemi et al.]
{C. S. Buemi$^{1}$ \thanks{E-mail: cbuemi@oact.inaf.it},
C. Trigilio$^{1}$,
P. Leto$^{1}$,
G. Umana$^{1}$,
A.~Ingallinera$^{1}$,
F. Cavallaro$^{1,2,3}$,
\newauthor L. Cerrigone$^{4}$,
C.~Agliozzo$^{5,6}$,
F.~Bufano$^{1}$,
S.~Riggi$^{1}$,
S.~Molinari$^{7}$,
F. Schillir\`o$^{1}$
\\
$^{1}$INAF - Osservatorio Astrofisico di Catania, Via S. Sofia 78, 95123 Catania, Italy\\
$^2$CSIRO Astronomy and Space Science, PO Box 76, Epping, NSW 1710, Australia\\
$^3$Universit\`a di Catania, Dipartimento di Fisica e Astronomia, Via Santa Sofia, 64, 95123 Catania, Italy\\
$^4$ASTRON, the Netherlands Institute for Radioastronomy, PO Box 2, 7990 AA Dwingeloo,The Netherlands\\
$^5$Millennium Institute of Astrophysics, Santiago 7500011, Chile\\
$^6$Universidad Andr\'es Bello, Avda. Republica 252, Santiago 8320000, Chile\\
$^7$INAF-Istituto di Astrofisica e Planetologia Spaziale, via Fosso del Cavaliere 100, 00133 Roma, Italy
}
\begin{document}

\date{}

\pagerange{\pageref{firstpage}--\pageref{lastpage}} \pubyear{}

\maketitle

\label{firstpage}

\begin{abstract}
We present a multi-wavelength study of the Galactic Luminous Blue Variable HR~Carinae, based on new high resolution mid-infrared (IR) and radio images obtained with the Very Large Telescope (VLT) and the Australia Telescope Compact Array (ATCA), which have been complemented by far-infrared \textit{Herschel}-PACS observations and ATCA archive data. The \textit{Herschel} images reveal the large-scale distribution of the dusty emitting nebula, which extends mainly  to the North-East direction, up to 70~arcsec from the central star, and is oriented along the direction of the space motion of the star. In the mid-infrared images, the brightness distribution is characterized by two arc-shaped structures, tracing an inner envelope  surrounding the central star more closely.  At radio wavelengths, the ionized gas emission lies on the opposite side of the cold dust with respect to the position of the star, as if the ionized front  was confined by the surrounding medium in the North-South direction. The comparison with previous data indicates significant changes in the radio nebula morphology and in the mass-loss rate from the central star, which has increased from 6.1$\times10^{-6}$~M$_{\odot}$~yr$^{-1}$ in 1994-1995 to $1.17\times10^{-5}$~M$_{\odot}$~yr$^{-1}$ in 2014. We investigate possible scenarios that could  have generated the complex circumstellar environment revealed by our multi-wavelength data.
\end{abstract}


\section{Introduction}
In the traditional view, Luminous blue variables (LBVs) are a class of massive and very luminous stars in the post main-sequence evolution stage, moving from the hot supergiant towards the Wolf-Rayet (WR) phase \citep{HD94,LN02}. Currently, their link to other advanced evolutionary phases of massive stars, such as supernovae, represents a hot topic of discussion \citep{Barlow_05, Kotak_06}. They are among the rarest types of stars because of the relatively shortness of this evolutionary phase, believed to last about 10$^4$-10$^5$ yr, and the high initial masses, greater than  22~M$_{\odot}$ \citep{Lamers_01}. Despite heir rarity, LBVs may play an important role in the chemical enrichment of the interstellar medium, due to the large amount of processed material injected by means of continuous winds and intense eruptions. During their short lives, LBVs undergo heavy episodic mass-loss events during which they lose a large fraction of their H-rich outer layers, as evidenced by the observations of their complex circumstellar environments.
The study of the geometrical and physical properties of these stellar ejecta could provide us with important insights on the mass-loss phenomenon and the evolution of the progenitor star.
The richness of their morphological structures suggests that the mechanisms that create and shape the extended nebulae are several and still poorly understood. Most proposed models are related to hydrodynamic explosions and continuum-driven super-Eddington winds, rather than line-driven stellar winds \citep{HD94, smow_06, smith_14}. There is a growing evidence for bipolar morphology occurring frequently in LBV nebulae \citep{White_00, Weis_01, Umana_05}, and several models have been proposed to reproduce such structures, most of them considering the shells as the result  of the interaction between a spherical symmetric stellar wind and an equatorial disc \citep{Nota_95, Ohara_03}.  However, it is unclear whether the bipolar nature of the LBV ejecta is due to a pre-existing density contrast or the wind itself is asymmetric \citep{Frank_98, Maeder_01}. For LBVs with ring nebulae, equatorial mass-loss (enhanced by the fast rotation of the star, the presence of a companion star, or a magnetic field) is often proposed \citep[see][and references therein]{2015Gvaramadze}.

Recently, the \textit{Herschel} and \textit{Spitzer} space missions allowed for a very detailed mapping of the circumstellar dusty environments associated with Galactic LBVs. At the same time, high angular resolution radio observations offer the additional possibility to reveal the ionized gas inside the dusty envelope. This provides useful clues to understand the interaction between the ejecta and the surrounding medium. The synergistic use of different techniques at different wavelengths is necessary for a good understanding of the physical conditions in LBVs, allowing us to analyse the different emitting components that coexist in the stellar ejecta. In some cases, evidence for asymmetric mass loss \citep{Buemi_10} and possible mutual interaction between gas and dust components \citep{Umana_10} has emerged from the comparison of the mid-IR and radio images of the nebulae, while important information on the gas and dust composition has been derived from the \textit{Spitzer}  and \textit{Herschel} spectra \citep{umana_09, Vamvatira-Nakou_13, Vamvatira-Nakou_15}.

HR~Car is a member of the small group of confirmed galactic  LBVs, showing the typical characteristic of photometric and spectroscopic S Doradus type variability \citep{Genderen_01}. The central star, classified as B2I, has a luminosity of 5$\times$10$^5$~L$_{\odot}$, as derived from spectral analysis of the Balmer lines performed by \cite{Machado_02}, and is located at a distance of 5.4$\pm$0.1 kpc \citep{Genderen_91}. The nebula associated with the central object, firstly reported by  \cite{Hutsemekers}, has a bipolar and filamentary structure, as evidenced by the coronographic imaging performed by  \cite{Clampin_95}, and  subsequently confirmed by   \cite{Weis_97} and  \cite{Nota_97}, who pointed out the similarity to the $\eta$~Car nebula. In particular, \cite{Nota_97} have suggested that the structure of the HR~Car nebula emerges from two bipolar lobes, expanding with a maximum velocity of about 115~km~s$^{-1}$. The mid-IR images obtained by \cite{Voors_97} reveal that, at arcsec scales, the geometry of the inner nebula around HR~Car is not point symmetric with respect to the central star and shows a morphological mismatch with the large-scale structure and the ionized-gas distribution mapped at 12.8~$\mu$m in the narrow-band [NeII] filter. The authors concluded that the nebula is composed of multiple shells resulting from different mass-loss episodes, probably occurred with a time-dependent geometry. The same conclusion was reached by \cite{White_00} based on radio observations, which indicated a strongly asymmetric inner nebula and an extended emission, whose morphology closely resembles that of the H$\alpha$ bipolar nebula.  The author also suggested the hypothesis that the presence of a companion  might account for the asymmetry in the nebula. On the basis of interferometric 
measurements performed with the Very Large Telescope Interferometer (VLTI) at different epochs,  \cite{Boffin_16}
have shown that HR Car is a binary system. The authors derived an approximate orbital period between 
a few to several tens of years, but they ruled out that the detected companion can be the object suggested by \cite{White_00}.
\begin{figure*}
\includegraphics[scale=0.9,bb=0 250 571 518,clip]{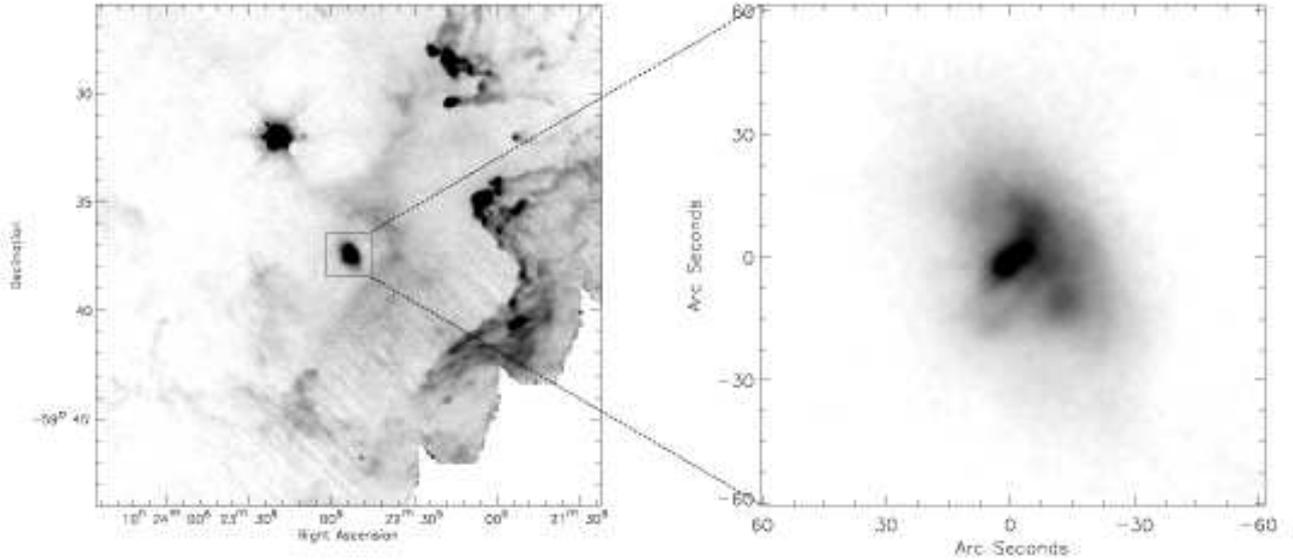}
\caption{The image of the environment around HR Car, provided by Hi-GAL at 70~$\umu$m, along with the more detailed view of the circumstellar environment from MESS at the same wavelength.}
\label{hr_pacs}
\end{figure*}
A detailed mid-IR spectroscopic investigation was performed by  \cite{umana_09}, who reported for the first time the existence of photodissociation regions (PDR) in the nebulae surrounding LBV objects and observed an abundance enhancement of Fe in the gas phase, which is considered as indirect evidence of shocks occurring in the nebula. The presence of amorphous silicates in the dusty circumstellar medium, together with the lack of crystalline silicates in the inner part of the nebula, suggests that recent dust formation occurred in HR Car during the LBV eruptions.

In this paper, we present the results of a multi-wavelength study of the complex circumstellar environment associated with the LBV HR~Car. The analysis is based on a new set of mid-IR, far-IR, and radio images obtained with the \textit{Herschel} Space Observatory, the Very Large Telescope (VLT), and the Australia Telescope Compact Array (ATCA). In Section 2, we describe the observations and data reduction. 
In Sect. 3 and 4, we discuss the analysis of the images obtained from each data set, 
probing different components of the nebula. 
In Sect. 5, we compare the brightness distribution and the asymmetries at
the various wavelengths, with the aim to obtain a comprehensive overview of the envelope 
and thus explain the observed characteristics. Finally, a summary of the main
conclusions is given in Sect. 6

\begin{table}
\begin{center}
\caption{VLT/VISIR Observations Log}
\begin{tabular}{ccccccc}
\hline
Filter     &  $\lambda_c$ & Date      &  Integration  & Airmass & FWHM    \\
           &   ($\umu$m)   &        &    time (s)              &         & (arcsec)    \\
\hline
SIV       &  10.49 &2011 Feb 23 & 1200  & 1.35 & 0.43\\
PAH2\_2  & 11.88 &  2010 Dec 14 & 2000  & 1.37& 0.46\\
\hline
\end{tabular}
\label{visir_log}
\end{center}
\end{table}

\section{The data sets}
\subsection{\textit{Herschel}/PACS imaging}
HR~Car was observed as part of both the Hi-GAL \citep{Molinari_10} key project on 2012 November 14 and the Mass loss of Evolved StarS (MESS) key project \citep{Groenewegen_11} on 2010 August 16, using the Photodetector Array Camera and Spectrometer (PACS, \citealp{Poglitsch_10}) on board of the {\it Herschel Space Observatory}.  The observations were performed by using different observing modes because of the different scientific rationales. Hi-GAL maps were obtained with the `fast scan parallel' mode and the final images were generated with pixel sizes of 3.2 and 4.5~arcsec at 70 and 160~$\umu$m respectively \citep{Elia_13}. The Hi-GAL pipeline described in \cite{Traficante_11} was used for the data reduction.

PACS MESS observations were performed using the `scanmap' mode at 70, 100 and 160~$\umu$m. As described by  \cite{Groenewegen_11}, the images were oversampled by a factor of 3.2 and characterized by pixel sizes of 1~arcsec at 70 and 100~$\umu$m and 2~arcsec at 160~$\umu$m. More details about the data reduction can be found in \cite{Ottensamer_11}. Data analysis was performed on the `Level 2' data products retrieved from the {\textit{Herschel} Science Archive and no additional data reduction procedure was performed.

\subsection{VLT/VISIR}

Mid-infrared imaging observations of HR~Car were conducted in service mode using the VLT Imager and Spectrometer for the mid InfraRed (VISIR, \citealp{Lagage}), mounted on the Cassegrain focus of the VLT Unit Telescope 3 (Melipal).  The imaging was carried out through the filters SIV (10.49/0.16 $\umu$m) and PAH2 (11.88/0.37 $\umu$m) on 2011 February 23 and 2010 December 14 respectively. The 0.127~arcsec pixel scale was used, corresponding to a 32$\farcs$5 $\times$ 32$\farcs$5 field of view. The standard chopping/nodding technique was applied to remove the sky contribution, with chopping and nodding throw angles of 20~arcsec and a chopping frequency of 0.25~Hz. To further improve the image quality, a random jitter pattern with a maximum throw of 3~arcsec was superimposed on the nodding sequence. The standard stars HD89682 and HD152880 were observed and used for photometric calibration in both bands. Such observations were also used to determine the FWHM of standard star images and thus to derive the actual angular resolution of our final maps. A summary of the VISIR observations is presented in Table~\ref{visir_log}, including the date of the observations, the exposure time, the airmass at the beginning of the observations, and the FWHM. 

The images were processed using the standard pipeline for data reduction provided by ESO (version 3.3.1). The chopped and nodded images were then combined to make a single image for each filter.

\begin{table}
\begin{center}
\caption{Log of the ATCA archive observations}
\begin{tabular}{cllcc}
\hline
Project Name     &   Date      &  Frequency   & Array     \\
          &             &       (GHz)             &          \\
\hline
C312   &  1994 Aug 22/23 & 8.6  & 6A  & \\
C186   &  1994 Sept 04/05 & 8.6  & 6A &\\
C312   &  1995 Jan 02     & 9.2  & 6A &\\
C312  &  1995 Apr 11     & 8.3, 9.0  & 6C &\\
C186   &  1995 Nov 08/09 & 8.6  & 6A &\\
C1167   &  2004 July 15 & 18  & 6A &\\
C2943   &  2014 Dec 24 & 5.4, 9.0, 44.0  & 6A &\\

\hline
\end{tabular}
\label{atca_log}
\end{center}
\end{table}

\begin{figure}
\includegraphics[scale=0.9]{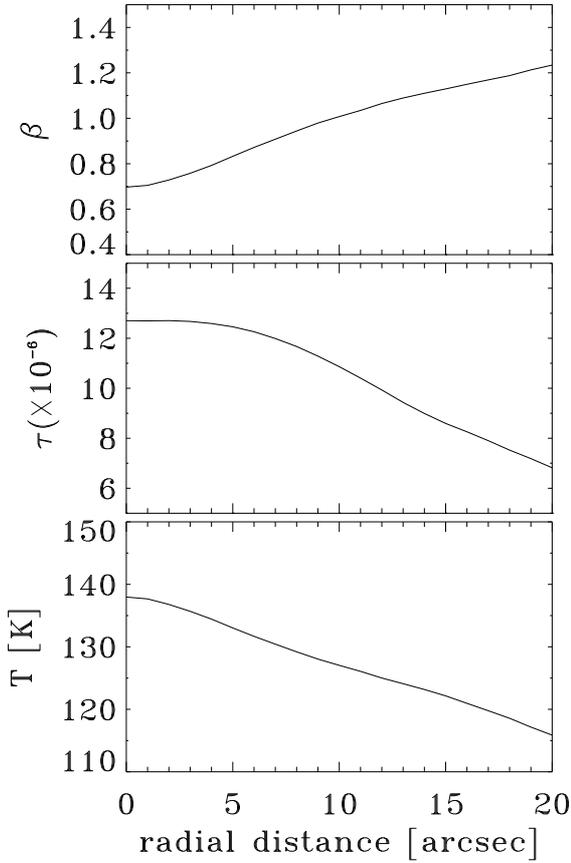}
\caption{ {Average radial profiles of the dust temperature (T), opacity index ($\beta$), and optical depth ($\tau_{0}$) maps, as obtained by a pixel to pixel fit of the SED by a modified single-temperature blackbody (see text).}}
\label{slices}
\end{figure}

\subsection{ATCA}
We observed HR~Car with the ATCA in 2014, on 24 December, with the array in 6A configuration, using the Compact Array Broadband Backend (CABB), which provides two independent intermediate-frequency (IF) outputs of 2 GHz in dual linear polarization. The data were taken interleaving simultaneous observations at two frequency pairs: 5.5/9~GHz and 43/45~GHz. All the observations used the 1M mode, which samples 2048 1-MHz wide channels. 

Bandpass and flux calibration were performed with observations of the standard calibrator 1934-638. For the phase calibration we used 1045-62. The beam size  of the observations was $1\farcs9\times1\farcs5$, $1\farcs3\times1\farcs1$, and $0\farcs7\times0\farcs2$ for the 5.5, 9, and 44~GHz observations, respectively. 

We used the  \textsc{miriad} data reduction package to process all data sets. 
We split the CABB \textit{4~cm} band in two
2-GHz sub-bands, one centred at 5.5~GHz and the other at 9 GHz, then we calibrated and imaged the data across the entire 2-GHz bands. For the data set obtained with the \textit{7~mm} receiver, we merged the two bands centred at 43 and 45~GHz after the calibration process  to obtain an average map with a high signal-to-noise ratio. In all these cases, the broad-band data were weighted with the robust parameter setted at 0.5 and multi-frequency synthesis was applied, using the \textsc{miriad} task INVERT. The images were subsequently deconvolved with the CLEAN algorithm, as implemented in the task \textsc{mfclean}, and finally restored with a synthesised beam (task RESTOR).

\subsubsection{Archival Data}
\label{archive}

The Australia Telescope Online Archive (ATOA) was queried for observations of HR~Car. In addition to project C312 and C186 \citep{White_00, Duncan_02}, project C1167 targeted HR~Car at 18~GHz. We selected the observations performed with 6~km array configurations, to obtain homogeneous data sets for map comparison. Table~\ref{atca_log} summarises the dates and observing frequencies of these observations. For a better comparison with our data set, these observations were processed following the same scheme described in the previous section. All the observations performed at the same frequency were concatenated in a stacked data set to produce a single map. The achieved spatial resolution (FWHM) and rms noise level are 1.2$\times$1.2~arcsec$^2$ and 2.4$\times$10$^{-5}$ Jy~beam$^{-1}$ at 9~GHz and 1.1$\times$0.8~arcsec$^2$ and 7$\times$10$^{-5}$ Jy~beam$^{-1}$ at 18~GHz.

\section{Shape and structure of the dusty nebula}

\subsection{The extended nebula}

In Figure~\ref{hr_pacs}, we present the complex environment around HR~Car as found in the 70~$\umu$m Hi-GAL PACS image. It shows extended dusty clouds and very clear complexes of fingers of emission in the West side. The outermost and brightest structures are probably associated with HD~302821, as  the pillars point toward the supergiant located about 6~arcmin away from HR~Car in the NE direction. On the other hand, the fainter arc-shaped emission at the western side of the HR~Car nebula is located at the same distance of HR~Car and is probably interstellar material that has been swept up by the stellar wind \citep{Nota_97}.

Together with the higher-resolution 70- and 100-$\umu$m images from MESS, these maps provide us with  valuable information on the structure of the dusty nebula, revealing a circumstellar emission much more extended and complex than previously known. The presence of a cold dust component in the circumstellar environment of HR~Car was suggested by \cite{Lamers_96}, to account for the jump observed in the  spectra taken in different bands with the \textit{ISO-Short Wavelength Spectrometer}.
\begin{figure}
\rotatebox{-90}{\includegraphics[scale=0.8,angle=90]{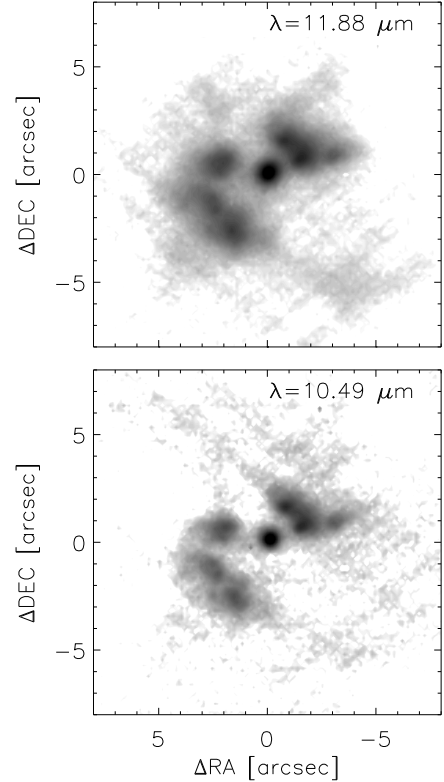}}
\caption{HR~Car in the PAH2\_2 and SIV filters with VISIR camera. North is up, and East points to the left.}
\label{hr_2vlt}
\end{figure}

The nebula has an essentially elliptical shape, with the major axis oriented almost along the NE-SW direction. Its size is about 50$\arcsec\times$70$\arcsec$, that is about 1$\times$1.5~pc at 5~kpc (P.A. $\sim$30$^\circ$ North to East) and has an integrated flux density at 70~$\umu$m of about 23.65~Jy. \par 
The most noticeable characteristic is the distribution of the emitting material, which extends mainly to the NW side with respect to the central star, suggesting a high degree of asymmetry in the material lost by the star. The comparison of the \textit{Herschel}  image with the H${\alpha}$ one (Fig.~1 in \citealp{Nota_97}) shows that the brightest part of the infrared emission correlates with the H${\alpha}$ features, resembling the bipolar lobes lying in the SEÐNW direction. The  H${\alpha}$ Southern lobe is also clearly noticeable, but it is evident that the dusty envelope extends well beyond the ionized-gas component, especially to the North-East side of the nebula.
\begin{figure}
{\includegraphics[scale=0.65]{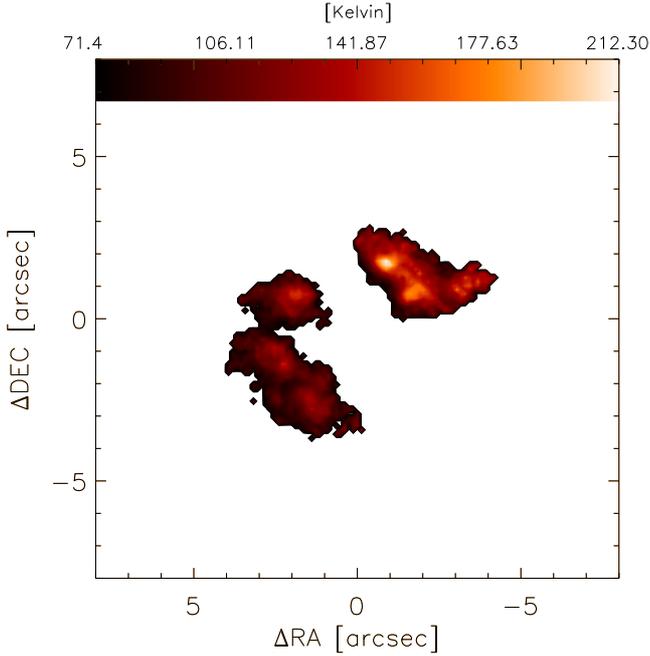}}
\caption{Temperature (colour) map derived by combining the images at 10.49 and 11.88~$\umu$m.}
\label{temp_visir}
\end{figure}
Driven by the structured dust-envelope appearance, we focused on the analysis  of the Spectral Energy Distribution (SED) across the nebula, with the aim of deriving some hints of possible variations in the dust characteristics. For each pixel of the 70-$\umu$m map, a SED was thus constructed using the MESS 70-, 100-, and 160-$\umu$m maps and the 12- and 22-$\umu$m from Wide-field Infrared Survey Explorer (WISE; \citealp{wright_10}) images. All the maps were convolved with the same beam of 16.8 arcsec (that is the FWHM of the WISE 22-$\mu$m co-added images) and then regridded to correspond the pixel-size with the 70-$\umu$m images. The SED of each pixel was then fitted by a single-temperature blackbody modified by an opacity index ($\beta$). Assuming that the flux density $F_{\nu}$ at frequency $\nu$ is given by 
\begin{displaymath}
F_{\nu}\propto B_{\nu}(T)\times(1-e^{-\tau_{\nu}}),~~~\mathrm{where}~~~\tau_{\nu}=\tau_0\times (\nu/\nu_0)^{\beta}
\end{displaymath}
where $B_{\nu}$ and $\tau_{\nu}$ are the Planck intensity and the optical depth at frequency $\nu$, from the fit we obtained the dust temperature $T$, the opacity $\beta$ and the optical depth $\tau_{0}$ at the chosen reference wavelength ($c/\nu_0=100$~$\mu$m). The \textsc{idl} code MPFITFUN based on a least-square $\chi^2$ fit has been used. In Figure~\ref{slices} the average radial profiles for the resulting  $T$, $\beta$, and $\tau_{0}$ maps 
starting from the central star position are shown. The profiles suggest the presence of a hot dust region at distance smaller than about 8~arcsec from the central object, where the optical depth is higher, and of a more extended colder and optically thinner halo. Also the dust opacity shows a radial gradient, with increasing values of $\beta$ as we move away from the central core. Values of $\beta$ close to 2 at far-IR wavelengths are expected for standard small grains characterising the ISM, while smaller values of $\beta$ are usually attributed to larger grains, resulting from grain accretion. The radial dependence may indicate a change in the properties of the circumstellar dust, with more processed dust located in the inner region, where the eruptive episodes create the conditions favourable to the growth of dust grains \citep{Kochanek_11}, and smaller or more fragmented grains far from the star, where it is possible that the ejected material is mixed with pre-existing ISM.

The approximate dust mass of the emitting envelope can be derived from the optical depth map, by dividing it for the absorption coefficient per mass unit and then by integrating over all the pixels with brightness higher than 3$\sigma$ \citep{Umana_10}. The absorption coefficient for mass unity was computed using the package \textsc{OpacityTool} given by the DIANA project\footnote{http://www.diana-project.com/data-results-downloads/fortran-package-to-compute-diana-standard-dust-opacities/}. Assuming standard astronomical silicates, as suggested by the silicate emission bump observed by \cite{umana_09}, we derive a $\kappa_{100}=66.4$ $\mathrm{cm^{-2}\ g^{-1}}$ and a dust mass, within 70~arcsec, of $M_d=1.6~\times 10^{-3}$~M$_{\odot}$ (assuming a distance of 5.4 kpc).  As the infrared observations  just trace the hotter dust emission, such value has to be considered as a lower limit to the entire envelope dust mass content.
Assuming a dust-to-gas ratio of 200 for the circumstellar material, we can estimate a total nebular mass of at least 0.3~M$_{\odot}$.  This result is close to the estimate of 0.15~M$_{\odot}$ obtained by  \cite {Hutsemekers} assuming a bipolar cone geometry, and is consistent with the estimates of 0.8~M$_{\odot}$ and 2.1~M$_{\odot}$ derived by \cite{White_00} and \cite{Clampin_95}. Note, however, that in some cases the mass estimates could be affected by the uncertainty on the nebular geometry, which in HR~Car seems to be very complex.

\subsection{The inner dust envelope}
Through the mid-IR images of HR~Car (Fig.~\ref{hr_2vlt}), it is possible to investigate more deeply the inner dusty circumstellar envelope (CSE) and obtain valuable information about it. HR~Car was mapped  by \cite{Voors_97} in the mid-IR $N$ filter with the instrument TIMMI (Thermal Infrared Multi Mode Instrument). They reported about a quite compact dusty nebula, not symmetrically distributed around the central star. The better re\-so\-lut\-ion and the high sensitivity of the VLT allow us to discern finer details of the  nebular morphology.  
The two maps at 10.49 and 11.88~$\umu$m exhibit very similar features, with the emission mainly located along two arcs of different intensity, extending at least $\sim$5~arcsec South-East and 4~arcsec North-West of the central source.  In addition, the lowest flux level in the PAH2\_2 map reveals a very faint extended halo around the bright structures, visible out to 7~arcsec from the centre, which is not detected in the $SIV$ map.
Given the large extension of the dusty nebula revealed by the PACS images, it is important to note that if the cloud were more extended than the chop positions, then any large-scale emission - if present - would be cancelled out  in the mid-IR observations.

\begin{figure*}
\includegraphics[scale=0.36]{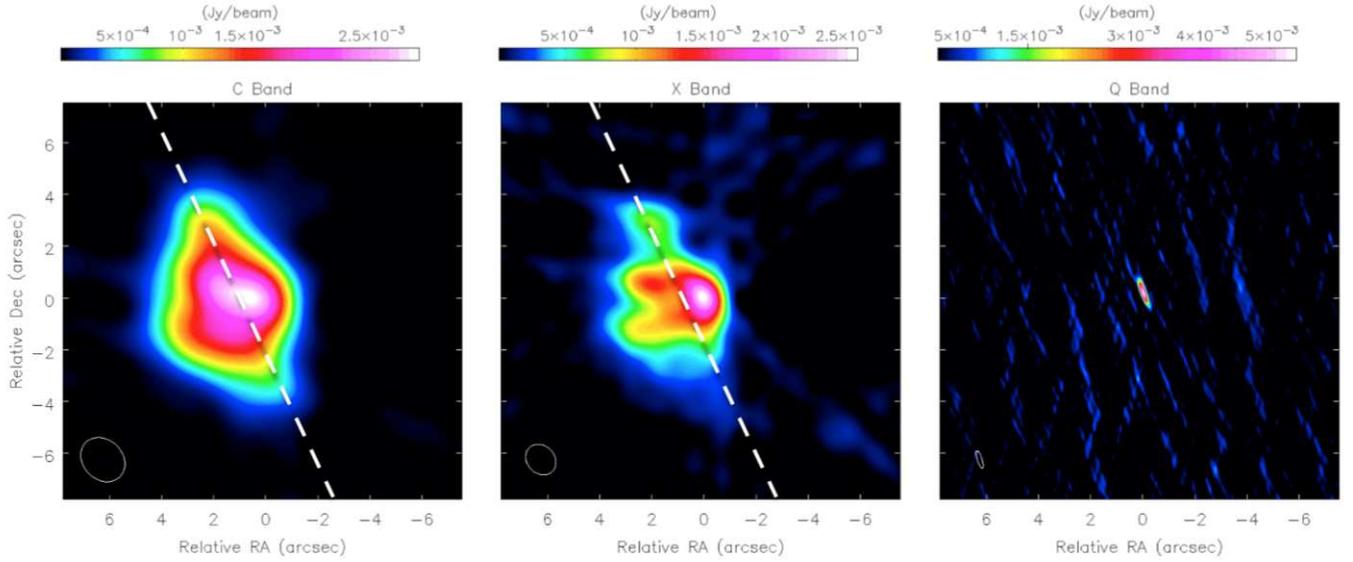}
\caption{The new ATCA radio images  of HR~Car at 5 (left), 9 (centre), and 44 (right) GHz. The dashed line indicates the major axis of the nebular elongation and is inclined of 25$^{\circ}$ with respect to the North-South direction.}
\label{newatca}
\end{figure*}

\begin{table}
\begin{center}
\caption{Properties of the new radio maps. }
\begin{tabular}{cccccc}

\hline
  Band  & Frequency&HPBW & LAS & Flux density & rms \\
   & [GHz]& [arcsec] & [arcsec] &      [mJy]   &     [mJy~beam$^{-1}$]             \\
  \hline
 C & 5.4&  1.9$\times$1.5 &  20.4  &       15.28       &  0.02 \\
 X &  9.0 & 1.3$\times$1.1 & 12.2  &       13.87      & 0.03\\
 Q  &  44.0~& 0.7$\times$0.2 &   2.5      &        5.9       & 0.2\\
  \hline
\end{tabular}

\end{center}
\label{pippo}
\end{table}

Under the assumption that the dust emission is optically thin at these wavelengths, by comparing the two images in the two bands, we derived the colour-temperature map displayed in Fig.~\ref{temp_visir}, as described in \citep{Umana_10}. The map shows a temperature gradient across the CSE and the resulting colour temperatures are in very good agreement with those obtained from the MESS maps. The higher spatial resolution of the VISIR images indicates that the NW arc is cooler than the SE crescent.

The optical-depth map at 10.49~$\umu$m was calculated from the relation $I_{\nu} \approx B_{\nu}(T) \tau_{\nu}$. The resulting dust mass (obtained as explained in the previous section) is $M_d=2.3 \times 10^{-4}$ M$_{\odot}$. This indicates that the mass of the hotter dust lying within 5~$\arcsec$ is about 14~percent of the mass of the entire dust envelope emitting in the far-IR.

\section{Radio emission}

Final images for radio bands at 5.5, 9, and 44~GHz are presented in Figure~\ref{newatca}. Table~\ref{pippo} reports the map properties and the measured fluxes, together with the array largest angular scale (LAS). The emission at 5.5 and 9~GHz traces in detail the structure of the extended ionized circumstellar nebula, which appears elongated $\approx$25$^{\circ}$ in the North to East direction. 
The resolution of the 44~GHz map allowed us to separate the stellar radio source from the extended ionized circumstellar nebula, filtered out at this frequency because of the small LAS.
The compact source is at $\alpha=10^\mathrm{h}22^\mathrm{m}53\fs9$, $\delta=-59\degr37\arcmin28\farcs3$ [J2000] and its flux density is equal to $F(44\ \mathrm{GHz})=5.9\pm0.2$~mJy.

Consistently with previous radio observations, the 5 and 9-GHz images reveal the morphology and the internal structure of the inner nebular core, showing a clear eastward displacement of the ionized region with respect to the peak of the emission, located at the position of the compact source observed at 44~GHz. The comparison of our maps with those obtained by \cite{Duncan_02} {points to some differences in the emitting features, despite the similarity in angular size and measured fluxes and the elongated elliptical morphology. In Figure~\ref{radio_95_15}, the 9-GHz map is displayed along with the 9 and 18-GHz maps obtained from the ATCA archive data sets. For a better comparison, we convolved and re-gridded all the maps to a common pixel size of 0.17~arcsec and resolution of 1.27$\times$1.07~arcsec$^2$. First of all, we notice that the 1995 map exhibits a main component coincident with the stellar object and a second clump at the eastern side of it, having a quite similar flux level. In our images, the bright emission at the central star position clearly outshines the eastward feature, which shows a horseshoe shape extending along two  arms oriented roughly horizontally, more evident in the 9-GHz image.

With the aim of investigating the mechanisms responsible for the radio emission, the observations at 5 and 9~GHz were used to generate the map of the radio spectral index $\alpha$, defined as $F_\nu\propto\nu^{\alpha}$. The images were obtained with the same parameters of the reduction tasks as for our own data. In order to reach the same angular resolution at the two frequencies, we restored both maps with Gaussian beams of the same sizes. Pixels whose values were below $5\sigma$ were discarded.
\begin{figure*}
\includegraphics[scale=0.9]{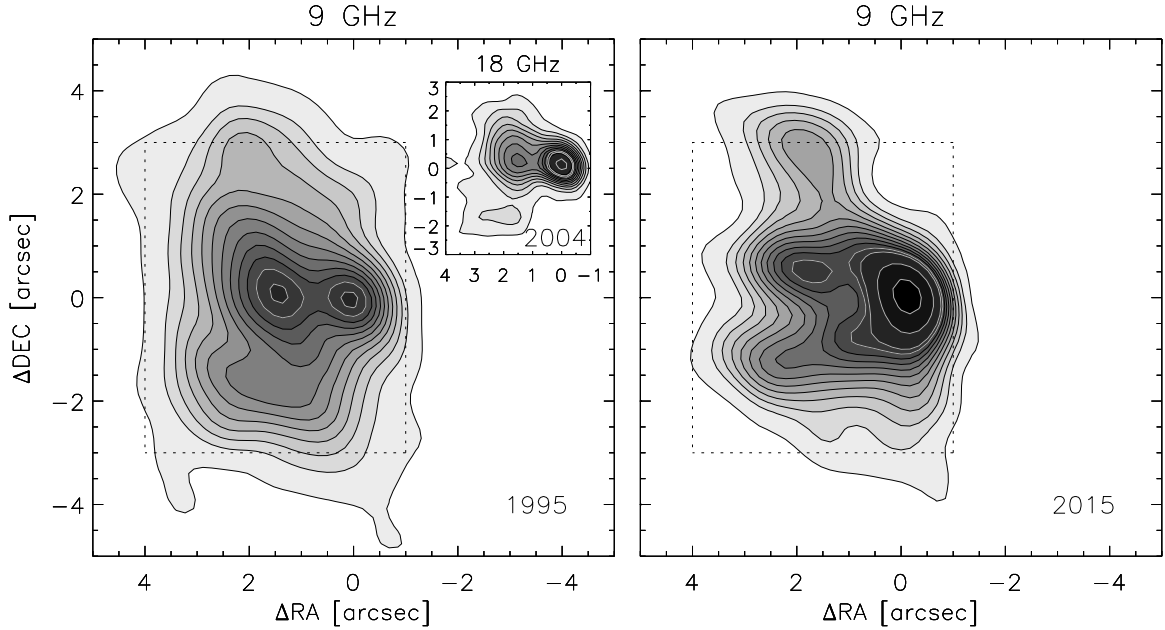}
\caption{VLA 9 GHz and 18 GHz images  at different epochs. The contours are drawn at the levels 1.5, 3, 4, 5, 6, 7, 8, 9, 10, 11, 12, 13, 17, and 23 of the rms in the maps, that is 1$\times$10$^{-4}$ mJy~beam$^{-1}$. For all the images, the half power beam width
(HPBW) of the synthesised beam is 1.27$\times$1.07~arcsec$^2$ and the pixel scales are set to 0.17 arcsec pixel$^{-1}$}. 
\label{radio_95_15}
\end{figure*}

\begin{figure}
\includegraphics[scale=0.65]{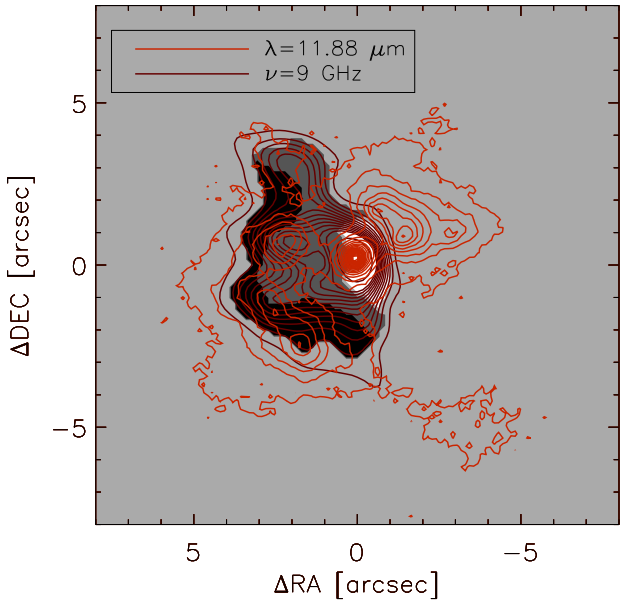}
\caption{Two frequency (5.5-9 GHz) spectral index intensity map with superimposed the 9-GHz (left panel) and the PAH2\_2 (right contour) map contours. The black area represents the region with $\alpha\leq-0.4$ (average $-0.6$), the grey area the region with $-0.4\leq \alpha \leq+0.2$ (average $-0.15$) and the white area the region with $\alpha \geq+0.2$ (average value $+0.45$). }
\label{spix}
\end{figure}
%
\begin{figure*}
\includegraphics[scale=0.9]{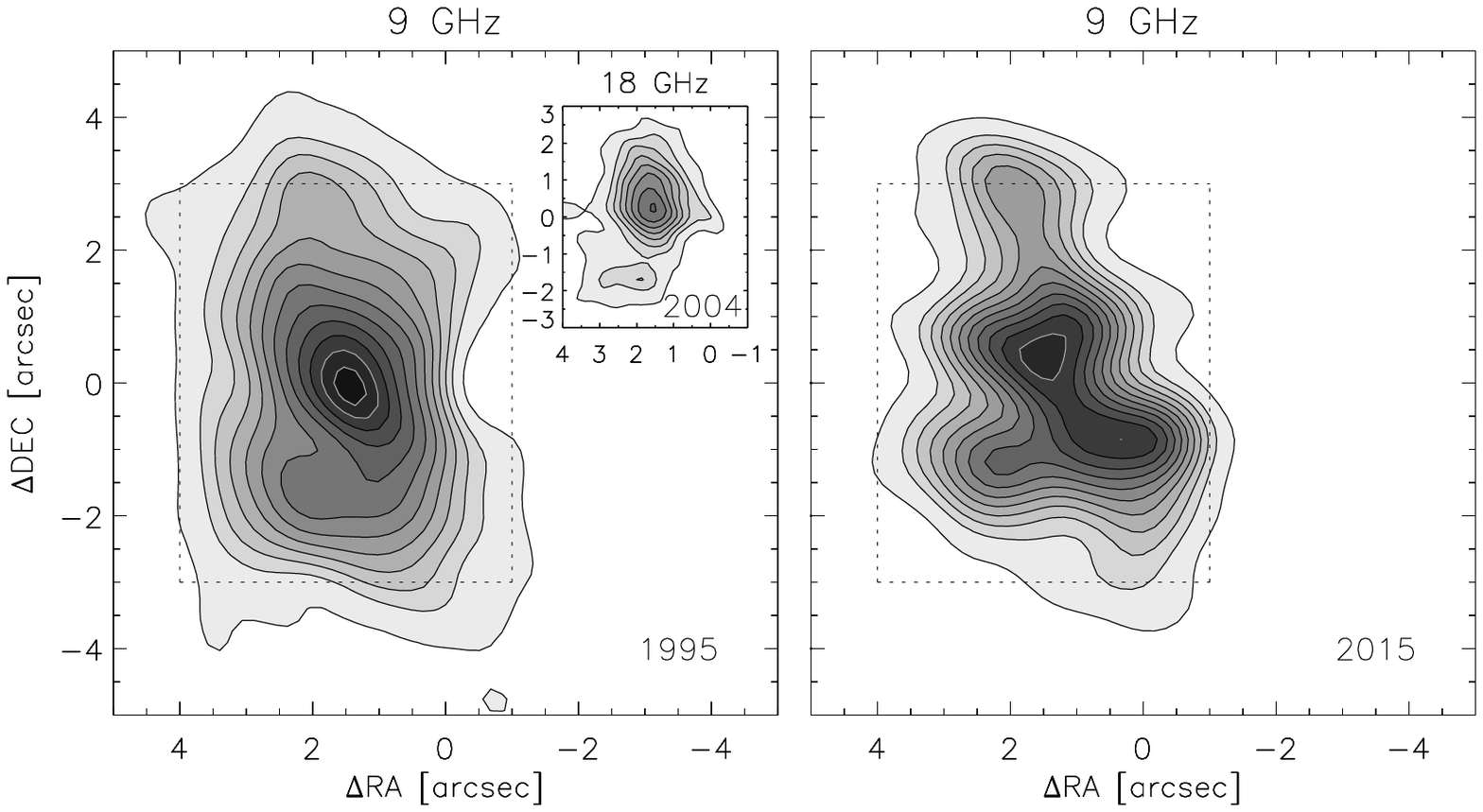}
\caption{The same maps as in Fig.~\ref{radio_95_15}, after the subtraction of the stellar component (see text).}
\label{residual}
\end{figure*}
%

In Figure~\ref{spix}, we present the spectral-index map and the contour of the same map superimposed on the VISIR PAH2\_2 image. It can be seen that the spectral-index distribution is  not uniform across the emitting region, indicating the presence of different components. At the stellar position, the spectral index is $\alpha=0.55$, indicating that in this region the radio emission is dominated by stellar-wind emission.   In the remaining part of the nebula, the spectral index ranges from values close to $-0.1$, typical of a free-free emission, to steeper values between $-0.4$ and $-0.7$ toward the edges of the radio-emitting region. Thus, though most of the diffuse emission is compatible with thermal emission, there are hints of a non-thermal emission component at the boundary of the brightest SE arm of the mid-IR emitting nebula (Fig.~\ref{spix}).

On the basis of the spectral-index map analysis, we assume that the flux measured at 44~GHz is due to thermal emission from the stellar wind. Therefore, the current mass-loss rate can be estimated from the relation derived by \citet{Panagia_75}:
\begin{center}
\begin{equation}
\centering
\dot{M}=6.7\times10^{-4}v_{\infty}F_{\nu}^{3/4}D^{3/2}(\nu \times g_{ff})^{-1/2}~[M_{\odot}~yr^{-1}],
\label{massloss}
\end{equation}
\end{center}
where $v_{\infty}=150$~km~s$^{-1}$ is the terminal velocity of the wind \citep{Hutsemekers}, $D=5.4$~kpc is the source distance \citep{Genderen_91} and $g_{ff}$ is the Gaunt factor, approximated with $g_{ff}=9.77(1 + 0.13$~log$\frac{T^{3/2}}{\nu})$ \citep{Leitherer_91}. Assuming a temperature of the plasma in the wind of 10$^4$~K,  we find that  in 2014 the star was losing mass at a rate of $1.17\times10^{-5}$~M$_{\odot}$~yr$^{-1}$.
\begin{figure*}
\includegraphics[scale=0.75]{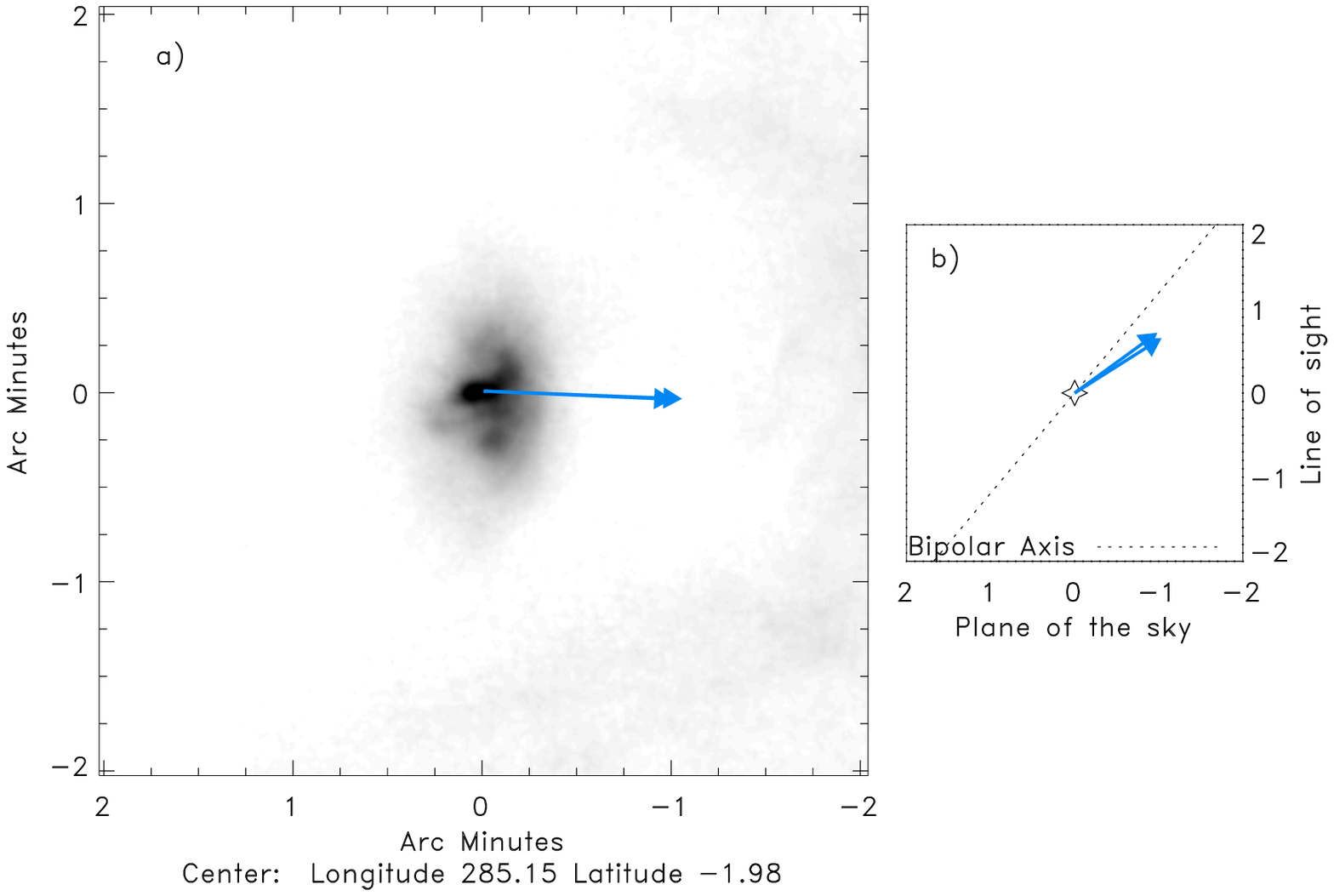}
\caption{Panel a) shows the PACS image of HR~Car, where the arrows indicate the direction of the space motion corrected for the solar motion. In panel b) the same arrows are sketched in a top view (90$^{\circ}$ out of the plane of the sky), to emphasise the angle between the direction of the space motion and the bipolar axis (dashed line), which is about 16$^{\circ}$.}
\label{hr_velocity}
\end{figure*}
The comparison between the new image at 9 GHz and earlier epoch maps (Fig.~\ref{radio_95_15}) highlights a flux density enhancement at the central star position. \cite{White_00} estimated, in fact, a total flux of 1.0~mJy at 9~GHz for the compact stellar component, against the 2.3~mJy derived by scaling the flux measured in 2014 at 44~GHz according to F$_{\nu}\propto\nu^{0.6}$. 
The mass loss rate derived by assuming  $\nu$~=~9~GHz and F$_{\nu}$~=~1~mJy in Eq.~\ref{massloss} is 6.1$\times10^{-6}$~M$_{\odot}$~yr$^{-1}$ between 1994 and 1995, that corresponds to a mass loss enhancement by a factor of $\sim$1.9. \par
HR~Car is known to be characterised by photometric variability, as shown by the data in the AAVSO website, and has brightened by almost two visual magnitudes over the past decade. During 1994 and 
1995 the source was in a phase of visual minimum of the S-Doradus cycle before it rose to the maximum, which began in 1996. In 2014 the star was again in the minimum phase, which is still on-going after the maximum ended in 2006. The visual magnitude at the minimum of 1995--1996 was 0.3--0.4 lower than in 2014, contrary to the lower mass-loss rate estimated. 
Such difference could be interpreted in terms of the bistability mechanism of line-driven winds, as a consequence of  a change in the iron ionization structure \citep{Vink_koter}, as it was hypothesised in the case of AG~Car \citep{Groh_11} and Iras~18576-0341 \citep{Buemi_10}.
On the other hand, the different mass-loss rate measured during the two minima could be related to the binariety of HR~Car, recently confirmed by  \cite{Boffin_16}, who constrained the time of the periastron passage between 2013.2 and 2015.05, and most likely in late 2014. The authors could not fully constrain the orbital period, but their best-fitting value corresponds to 12 years. This orbit parameters, if confirmed, locate the two stars much closer in 2014 than in 1994--1995, suggesting possible interactions between the stellar components that could affect the mass-loss rate near the periastron epoch.

To complement the study of flux variability, the ATCA archive was queried for observations with similar angular resolution, but no additional data were found on the time-line between the two data sets, with the exception of the 18~GHz observations performed in 2004 July, not previously published. The resulting map (Fig.~\ref{radio_95_15}) shows a doubly-peaked structure similar to the 1995 image, in excellent agreement with the flux from \cite{White_00} under the assumption of spherical stellar-wind emission. Together with the similarity in the structure of the nebula, this indicates that the measured flux variation took place after 2004, but the large temporal gap between the observations makes it impossible to rule out that further changes occurred.

To separate the nebular and the stellar-wind emission, a point source component at the stellar position has been subtracted from the data sets in the Fourier plane, via the \textsc{miriad} task UVMODEL, where the assumed model flux density was  1.0 and 2.3~mJy for the \cite{Duncan_02} data set and ours, respectively. The maps of the residuals (Fig.~\ref{residual}) show that some differences are also present in the radio nebula. 
In particular, the main differences are in the inner part of the nebula, where the emission is arc-shaped. This morphology may derive from a different density distribution (e.g. from new mass injection) or from the shielding of the stellar ionizing radiation.

\section{Discussion}
Because of the different morphology of the nebula unveiled by the observations at various wavelengths, determining the actual structure of the circumstellar environment that surrounds HR~Car is a tricky task: it appears to be complex and probably the result of different mass-loss episodes. The great asymmetry in the large-scale structure of the nebula revealed by the far-IR images can be explained in terms of density variation in the cloud. The asymmetry might be the result of stellar evolution in a circumstellar and/or interstellar medium characterised by a radially decreasing density. However, it is difficult to assess if such inhomogeneity is pre-existing in the ISM around HR~Car or is a consequence of its mass-loss process.
\begin{figure*}
{\includegraphics[scale=0.85]{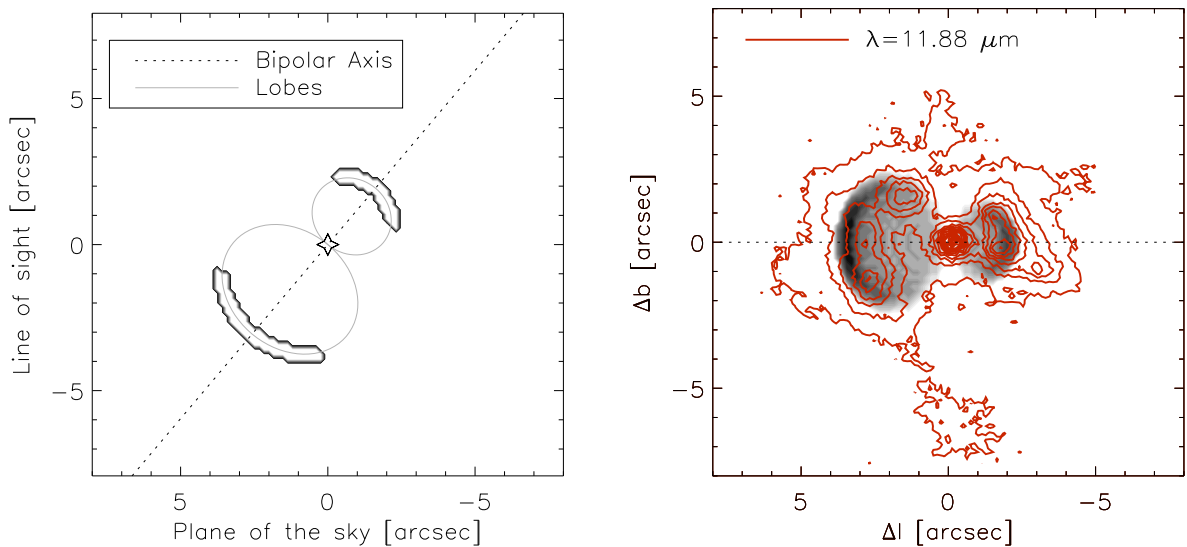}}
\caption{\ Schematic representation of the bipolar structure assumed in our simple geometrical model. The left panel sketches a top view of the two lobes originated from the star. The right panel shows the 11.88-$\umu$m (red lines) image contours overlapping the simulated emission from the model.}
\label{sketch}
\end{figure*}
We also explored the possibility that the arc-shaped morphology of the extended dusty nebula could have formed by compression in a bow-shock, driven by the stellar wind impacting on the surrounding medium, in the direction of the space motion of HR~Car.

Bow shock structures have been observed in the far-IR around different stellar objects, e.g. AGBs, red supergiants and Algols \citep{Ueta_08, Jorissen_11, Cox_12, Mayer_16}. First of all, we checked if the velocity direction of HR~Car is compatible with the location of the bow shock. The Galactic velocity of HR Car was determined following the calculation of  \cite{Johnson_87}. We assumed a proper motion of $\mu_\alpha=-6.87$ mas/yr and $\mu_\delta=2.33$ mas/yr \citep{Leeuwen_07} and a distance of $5.4$ kpc. Since \cite{Nota_97} gave two possible values of heliocentric radial velocity, $-14$ and $-22$ km sec$^{-1}$, we calculated the Galactic velocity for both cases. Taking into account the solar motion, in the first case we found $[U,V,W]=[-117.9,-29.5,-4.3]\ \mathrm{km~sec^{-1}},$ while in the second $[U,V,W]=[-120.0,-21.8,-4.0]\ \mathrm{km~sec^{-1}}$.

In both cases, the velocity vector lies almost perfectly on the Galactic plane with  angles of $2.0^\circ$ and $1.9^\circ$, respectively, which are very close to the direction of the line connecting the stellar position to the apex of the parabolic curve that defines the western boundary of the far-IR emission (Fig.~\ref{hr_velocity}).
A further issue with such scenario is the comparison of the nebular dimension with the expected wind stand-off distance, defined as the distance $r$ from the star where the momentum flux in the wind equals the ram pressure of the ambient medium. Under the assumption of a spherically-symmetric and time-invariable stellar wind and of a homogeneous ambient medium, the distance $r_0$ between the star and the apex of the bow (stagnation point) can be derived by the ram pressure balance of the wind and ambient medium:
\[
r_0= 0.18 \left( \frac{\dot{M}}{10^{-5}M_{\odot}~\mathrm{yr}^{-1}} \right)^\frac12  \left( \frac{v_w}{10~\mathrm{km~s}^{-1}} \right)^\frac12  \left( \frac{n_{AM}}{\mathrm{cm}^{-3}} \right)^{-\frac12} \times
\]
\begin{equation}
\times\left( \frac{v_{\star}}{100~\mathrm{km~s}^{-1}} \right)^{-1} \mathrm{pc},
\label{r0}
\end{equation}
where $v_w$ is the wind terminal velocity, $n_{AM}$ is the ambient medium density, and $v_{\star}$ is the velocity of the star \citep{Chiotellis_16}. The ambient density was derived using the relation given by \citep{Mihalas}, which relates the interstellar hydrogen nucleus density $n_\mathrm{H}$ to the distance from the Galactic plane $z$:
 \begin{math}
n_\mathrm{H}= 2.0~e^{-\frac{|z|}{100~pc}}~\mathrm{cm}^{-3}.
\label{nh}
\end{math}
Considering the 70-$\umu$m image, the star-apex distance projected on the sky plane is about 25~arcsec, which corresponds to 0.58~pc at 5.4~kpc. This distance is equal to 0.76~pc when de-projected following the system inclination with the respect to the plane of the sky proposed by \cite{Nota_97}. Adopting $v_{\star}$=120~km sec$^{-1}$ and $v_w$=115~km sec$^{-1}$ \citep{Nota_97}, from Eq.~(\ref{r0}) the stand-off distances of 0.76~pc is compatible with a bow shock caused by a mass loss rate of 6$\times$10$^{-6}$~M$_{\odot}$~yr$^{-1}$. Despite the rough approximations adopted, this value is in good accordance with the mass-loss rate typical of the LBV phase, as well as with the values derived for HR~Car on the basis of the radio data (Sec.~4).

It is noteworthy that the derived stellar velocity vector lies very close to the bipolar axis of the H$\alpha$ nebula (see panel b of Fig.~\ref{hr_velocity}). This suggests that the density enhancement produced in the bow shock may affect the distribution shape of the ejected material.

At shorter wavelengths, the VISIR images reveal a less extended emission  and a quite 
different inner nebular morphology, with brighter emission concentrated mostly in two 
arc-like structures that can be  interpreted as the results of the limb brightening of 
a circumstellar envelope close to the central star. 
The radio maps give us a detailed view of the innermost part of the ionized gas component of CSE, whose morphology adds one more piece to the puzzle of this complex environment. The radio images  show a marked asymmetry in the emission distribution with respect to the central object, but the asymmetry direction is opposite to that in the far-IR emission.

\subsection{The expanding-lobe model}
\cite{Voors_97} suggested 
that the CSE around HR~Car is the result of multiple outflows with a bipolar geometry.  This  was also
proposed by \cite{Nota_97} on the basis of  high-resolution coronographic observations.
The author interpreted the morphological and kinematic analysis of the HR~Car nebula as a large-scale bipolar structure, 
very reminiscent of the $\eta$~Car nebula, where the H${\alpha}$ filaments are the 
signature of two symmetrical expanding bubbles. The author hypothesised that the 
presence of bipolar lobes may imply an interaction between the fast stellar wind and a 
pre-existing CSE, in the form of an asymmetrical envelope or an equatorial dust density enhancement.

In the framework of the first scenario, a simple geometrical model of the two bubbles originating 
from the star is sketched in Figure~\ref{sketch}. The three-dimensional orientation of 
the structure is the same of the HR~Car H${\alpha}$ nebula derived by \cite{Nota_97}, 
with an inclination of the axis of the \lq\lq homunculus\rq\rq~to the plane of the sky  
of 50$^{\circ}$. Using de-projected radii of $2.5$ and 4~arcsec for the outer edges of 
the two expanding shells and $0.5$~arcsec for the shell thickness, the maxima in the 
optical depth, under the assumption of optically-thin conditions, overlap with the
maximum of the mid-IR intensity maps. The good agreement of the geometric parameters derived by \citet{Nota_97} with those used
 in our geometric toy model could indicate that there are no significant changes in the orientation of the outflows that create the structures at different epochs.\par
By a comparison of our mid-IR  and radio maps, the inner and bright ionized-gas emission is apparently confined within the dust and the warmer arc (see Fig.~\ref{temp_visir}) seems to trace the western boundary of the radio emission (Fig.~\ref{spix}). This suggests that the dust is optically thick to the UV photons, absorbing a portion of the stellar radiation and thus shielding the circumstellar gas from the ionizing flux. The overall elongation observed in lower-frequency radio maps could  indicate the direction in which the radiative flux is channelled by the surrounding medium. Apart from the stellar-wind emission, the spectral index appears to be flat ($\approx-0.1$) across most of the nebula, but steeper indices are found in the outer regions. Since the central star does not appear to produce a sufficient quantity of ionizing photons to keep the nebula ionized \citep{Nota_97} and considered the local variation of the radio spectral index, it is possible to hypothesise that a fraction of the radio-emitting electron population comes from the shock of the stellar wind interacting with the denser surrounding envelope. This could result in the gradually steepening of the spectral index toward the edge of the mid-IR emission structures, where $\alpha \approx-0.6$ matches with what expected from synchrotron emission due to shock-accelerated electrons.

\subsection{The jet-precession model}
\label{jet}

\begin{figure*}
\includegraphics[scale=0.55]{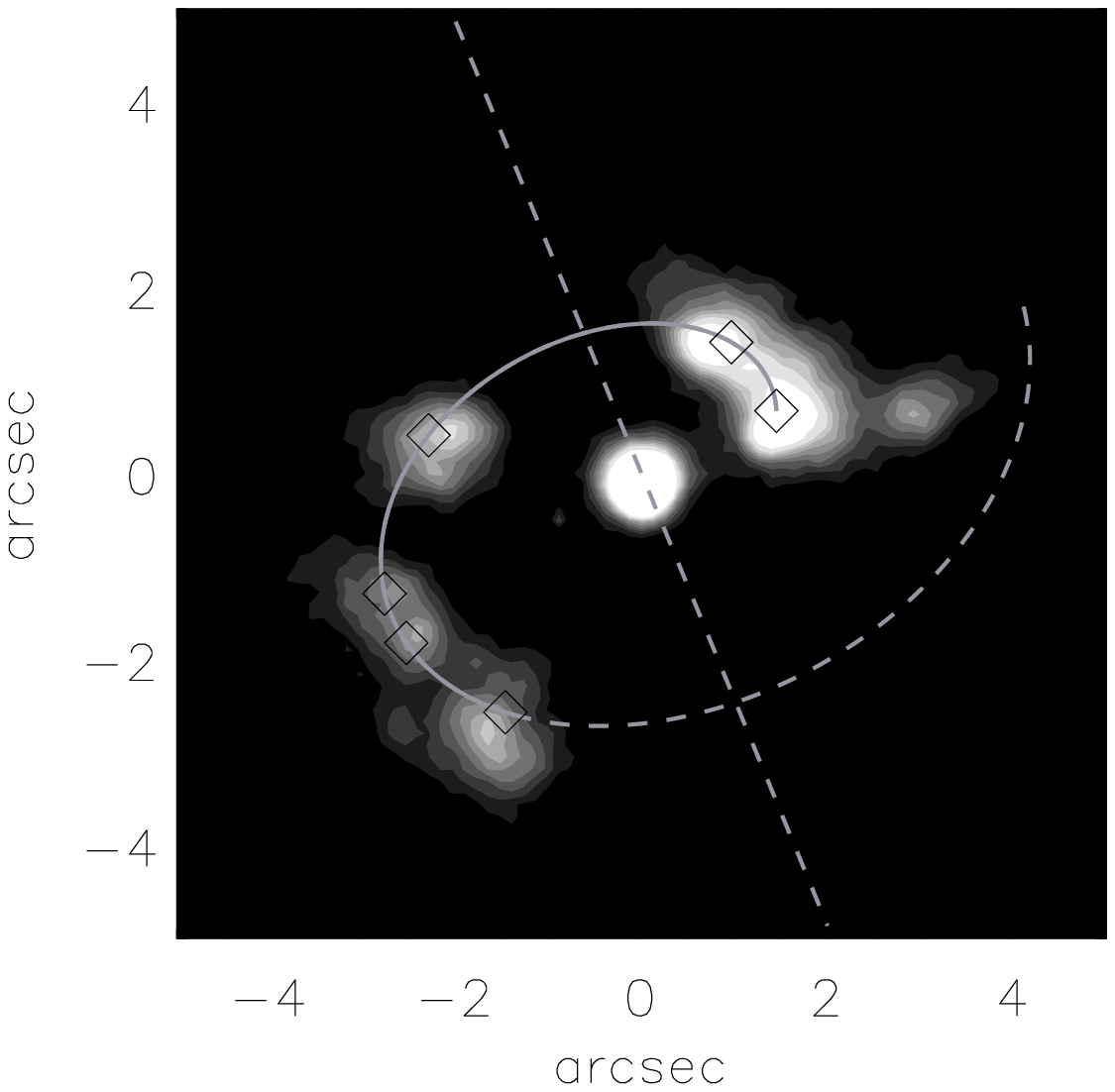}\includegraphics[scale=0.55]{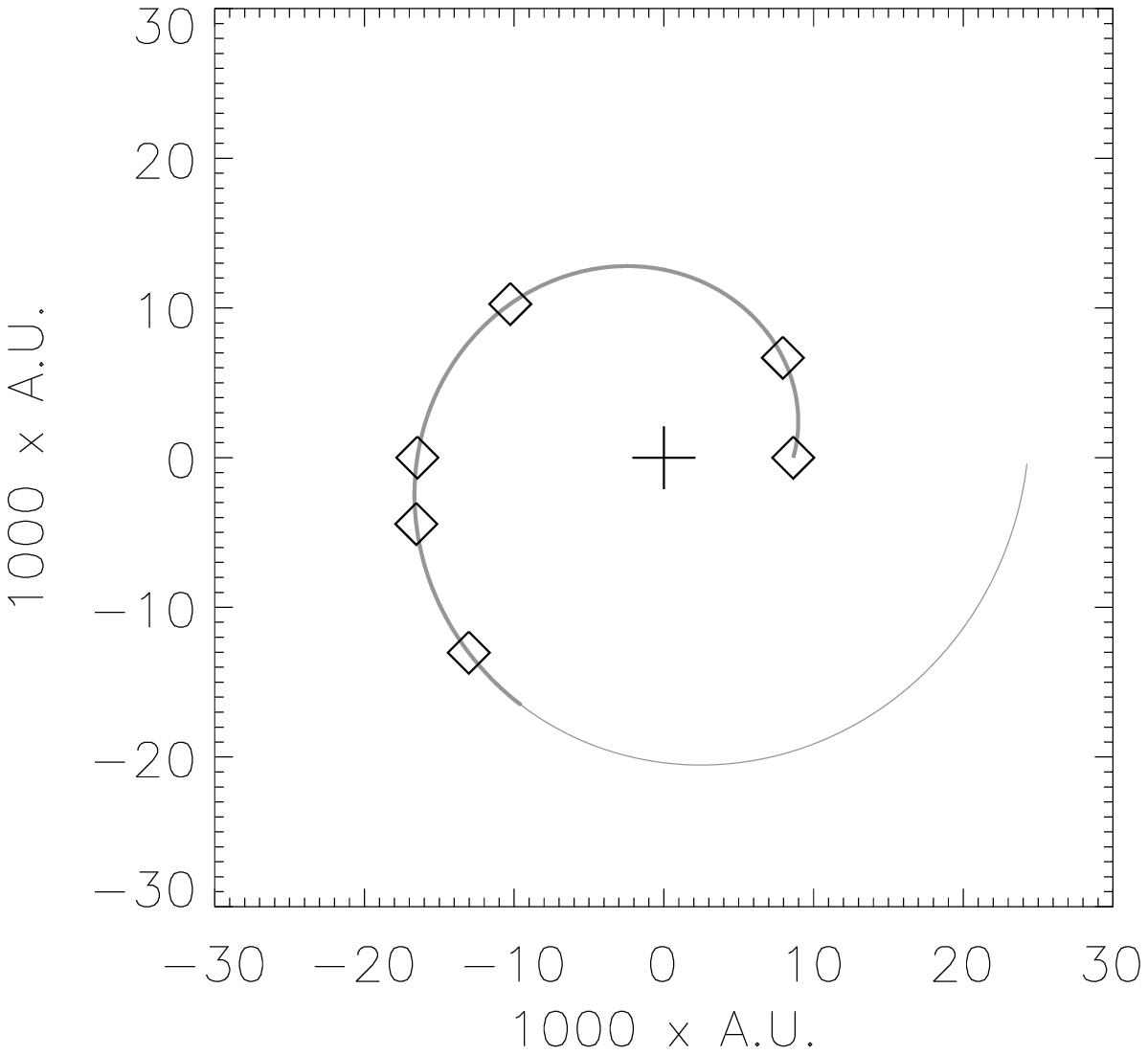}
\caption{Left panel: the mid-IR image of HR~Car at the SIV filter in the symbiotic model. The axis of the system (dashed line) is oriented 25$^{\circ}$ counter-clockwise. The blobs are located in an Archimedean spiral centred on the star. The plane where the spiral lies is inclined by 45$^{\circ}$ relatively to the line of sight. Right panel: the same spiral seen pole on; symbols locate the observed blobs; distances are in A.U. adopting a distance to HR~Car of 5.4~kpc}
\label{spirale}
\end{figure*}

\begin{figure}
\includegraphics[scale=0.25]{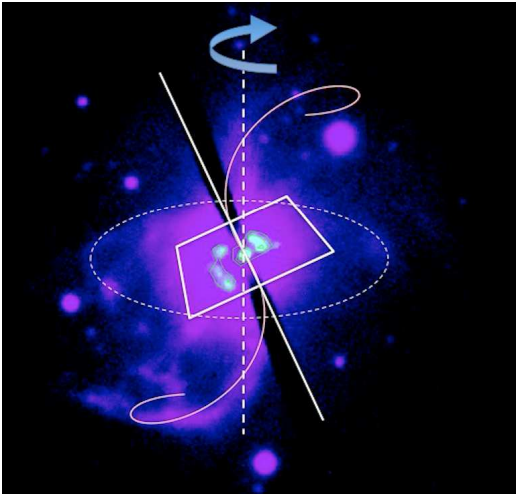}
\caption{Schematic view of the jet-precession model for HR~Car. The H$\alpha$ image from archive show the S-shaped jet. Mid-IR blobs form a spiral lying in a plane (represented by a square drawn in perspective). The outflow visible in radio and H$\alpha$ is perpendicular to this plane and precesses around an axis in North-South direction (dashed line). Diffuse H$\alpha$ emission is confined in the shades ellipse.}
\label{jet-precession}
\end{figure}

As shown in Figure~\ref{newatca}, the radio images at 5 and 9 GHz reveal that the ionized nebula is elongated, with a major axis at a position angle of about 25$^{\circ}$ (North to East). The mid-IR images at 10.49 and 11.88 $\umu$m show several blobs in an almost elliptical line around the central star, with the North-West features being closer to the star ($\sim$1.6~arcsec) than those in the South-East direction (more than 3~arcsec). The major axis of this pseudo-ellipse has a position angle of about 120$^{\circ}$ and the minor axis of about 25$^{\circ}$. It is worth to note that the elongation of the radio nebula and the minor axis of the pseudo-ellipse tracing the mid-IR features are almost identical. This suggests that the blobs visible in the mid-IR lie in a plane perpendicular to the axis of the radio nebula, in an almost circular region with centre offset with respect to the central star. A closer inspection of the mid-IR images reveals that the blobs are located in an Archimedean spiral centred on the star, where the distance $r$ to the blobs increases linearly with the polar angle $\theta$. Fig~\ref{spirale} illustrates the model. Adopting a distance of 5.4 kpc, the equation of the spiral is:
\begin{equation}
r=r_\mathrm{0}+\omega\times\theta
\end{equation}
with $r_\mathrm{0}=1.6$~arcsec, corresponding to 8640~A.U., or 0.042~pc, and $\omega=0.46$~arcsec/radians.

As already reported, the circumstellar features observed in H$\alpha$ have been interpreted by \cite{Nota_97} as a bipolar nebula with two lobes, with the symmetry axis oriented at an angle of $\approx 50^{\circ}$ (North to West). However, we note that these lobes are not complete, since the most evident features are an S-shaped emitting region centred on the star and a sort of elliptical diffuse emission elongated in the East-West direction (see Fig.~\ref{jet-precession}). The straight line tangential to the S-shaped feature at the position of the star has the same inclination of about 25$^{\circ}$ (North to East), coinciding with the major axis of the radio nebula and with the direction perpendicular to the plane where the \lq\lq spiral\rq\rq~lies. 

The idea is that of a jet-precession model, illustrated in Fig.~\ref{jet-precession}. It is still a preliminary model that will be developed in more details in a subsequent paper. The stellar wind is collimated along the rotational axis of HR~Car and gives rise to H$\alpha$ and Bremsstrahlung emission, the latter being due to the ionization caused also by a hot companion, the B0~V star supposed by \cite{White_00} in his symbiotic model. The presence of the hot companion in the East side can justify the values of the spectral index in the radio emission, as due to a wind-wind interaction. On the contrary, it is unlikely that the companion detected by \cite{Boffin_16} could contribute to the nebula ionization due to the low ionizing power (effective temperature) of red supergiants.\par
Violent episodes of copious mass loss occur in the equatorial plane when the stellar radius increases, approaching the equatorial rotational breakup velocity \citep{Groh_09}. These mass-loss events generate the blobs of the spiral visible in the near-IR. The direction of the ejection of the blobs may depend on the position of a companion star at the moment of the ejection. This star is probably either the B0~V or the closer massive red supergiant recently detected by \cite{Boffin_16} with the VLTI. \par
Although so far only one companion has been confirmed, if HR~Car is a triple system, it is subject to a precession that could give rise to a helical outflow observed as the S-shaped feature. The precession axis is approximately in the North-South direction. The matter ejected in the equatorial plane of the star over a time longer than the precession time accumulates in a circumstellar disc perpendicular to the precession axis. This is visible in the H$\alpha$ image inside the ellipse of Fig.~\ref{jet-precession}.\\

Note that hints for a jet-precession in the LBV star RMC127 in the Large Magellanic Cloud has been recently proposed by \cite{Agliozzo_16} and a precessing helical outflow has been observed in the Wolf-Rayet star WR102c by \cite{Lau_16}.

\section{Summary}
In this paper, we have presented a multi-wavelength study of the circumstellar environment surrounding HR~Car, based on new mid-IR data obtained with VISIR-VLT and new radio data obtained with the ATCA. Such observations have been  complemented by \textit{Herschel} images from the Hi-GAL and MESS projects and with radio ATCA archive data. We have found that the nebular morphology varies greatly with the wavelength, confirming the presence of a complex multiple-shell structure.

The \textit{Herschel} observations have revealed an extended envelope that reaches a distance of 70 arcsec from the central object toward NW. We derived  a dust mass content of at least 1.6 $\times 10^{-3}$~M$_{\odot}$ and, assuming a gas-to-dust ratio of 200, a lower limit of 0.3~M$_{\odot}$ for the total nebular mass, which is consistent with the values previously reported in the literature.
It is difficult to assess if the asymmetric distribution is the result of a density gradient in the ambient medium or it is caused by asymmetric mass-loss episodes that could be a consequence of binary interaction. We note that the location of the emitting structure and the derived dynamical age are consistent with a bow shock  resulting from the interaction of the circumstellar material with the ISM.

At mid-infrared wavelengths, the VISIR images show two arc-shaped structures surrounding 
the central star. These can be attributed either to the limb brightening of optically 
thin dust  in a bipolar lobe-like structure or to an equatorially enhanced mass loss such as a spiral outflow
arising from the binary interaction in the expanding wind of the mass-losing star. We discussed how the observed nebular morphology can be interpreted in the framework of the different proposed scenarios.

The new radio observations allowed us for the first time to clearly distinguish the central stellar wind from the nebular emission. We could then derive a mass-loss rate of 1.17$\times$10$^{-5}$~M$_{\odot}$~yr$^{-1}$ in 2015. This value is  1.9 times larger than the mass-loss rate estimated in 1994--1995. This variation does not seem to be correlated to the S~Dor cycle, but it could have been induced by the periastron passage of the companion star. \par
The brightest part of the ionized nebula lies on the opposite side of the bulk of the dust and seems to be well confined within the mid-IR  structure. 
Finally, the comparison of our new radio data with the   historic ATCA observations of HR~Car at 9 and 18~GHz, which we reduced in a coherent way with the rest of our data, revealed significant changes in both  the morphology of the nebula and the flux from the central star.

\section*{Acknowledgments}
This research is supported by ASI contract I/038/08/0 \lq\lq HI-GAL\rq\rq.
This work is based also on observations made
with the VISIR instrument on the ESO VLT telescope (programme
ID: 386.D-0727A).
The Australia Telescope Compact Array  is part of the Australia Telescope National Facility which is funded by the Australian Government for operation as a National Facility managed by CSIRO. This paper includes archived data obtained through the Australia Telescope Online Archive (http://atoa.atnf.csiro.au).\par
CA acknowledges support from FONDECYT grant No. 3150463.



\begin{thebibliography}{99}

\bibitem[\protect\citeauthoryear{Agliozzo et al.}{2016}]{Agliozzo_16}
   Agliozzo C. et al, 2016, submitted to ApJ
\bibitem[\protect\citeauthoryear{Barlow et al.}{2005}]{Barlow_05}
Barlow M. J. et al., 2005, ApJL, 627, L113
\bibitem[\protect\citeauthoryear{Boffin et al.}{2016}]{Boffin_16}
   Boffin H. M. J. et al., 2016, A\&A, 593, 90
\bibitem[\protect\citeauthoryear{Buemi et al.}{2010}]{Buemi_10}
Buemi C. S., Umana G., Trigilio C., Leto P., Hora J. L., 2010, ApJ, 721, 1404
\bibitem[\protect\citeauthoryear{Chiotellis et al.}{2016}]{Chiotellis_16}
Chiotellis A., Boumis, P., Nanouris N., Meaburn J., Dimitriadis G., 2016, MNRAS, 457, 9
\bibitem[\protect\citeauthoryear{Clampin et al.}{1995}]{Clampin_95}
Clampin M., Schulte-Ladbeck R. E., Nota A., Robberto M., Paresce F, Clayton G.C., 1995, AJ, 110, 251
\bibitem[\protect\citeauthoryear{Cox et al.}{2012}]{Cox_12}
Cox N. L. J., Kerschbaum F., van Marle A.-J., et al. 2012, A\&A, 537, A35
\bibitem[\protect\citeauthoryear{Duncan \& White}{2002}]{Duncan_02}
    Duncan R. A., White S. M., 2002, MNRAS, 330, 63
\bibitem[\protect\citeauthoryear{Elia et al.}{2013}]{Elia_13} 
     Elia D., et al., 2013, ApJ, 772, 45 
\bibitem[\protect\citeauthoryear{Frank et al.}{1998}]{Frank_98}
Frank A., Dongsu R., Davidson K, 1998, ApJ, 500, 291    
\bibitem[\protect\citeauthoryear{Gvaramadze et al.}{2015}]{2015Gvaramadze} 
 Gvaramadze V.~V., et al., 2015, MNRAS, 454, 219    
\bibitem[\protect\citeauthoryear{Groenewegen et al.}{2011}]{Groenewegen_11} 
     Groenewegen M.A.T., Waelkens C., Barlow M.J. et al., 2011, A\&A, 526, A162
\bibitem[\protect\citeauthoryear{Groh et al.}{1999}]{Groh_09}     
  Groh J.H., Hillier D.J., Barb\'a R., Fern\'andez-Laj\'us E., Gamen R.C., Mois\'es A.P., Solivella G., Teodoro M., 2009, ApJ, 705, L25
  \bibitem[\protect\citeauthoryear{Groh et al.}{1999}]{Groh_11}     
 Groh J.H., Hillier D.J. and Damineli A., 2011, ApJ, 736, 46
\bibitem[\protect\citeauthoryear{Johnson \& Soderblom}{1987}]{Johnson_87} 
     Johnson D. R. H., Soderblom D. R., 1987, AJ, 93, 864
\bibitem[\protect\citeauthoryear{Jorissen et al.}{2011}]{Jorissen_11}
Jorissen A., et al., 2011, A\&A, 532, 135
\bibitem[Humphreys \& Davidson(1994)]{HD94} Humphreys R.~M., Davidson K., 1994, PASP, 106, 1025
\bibitem[\protect\citeauthoryear{Hutsem\'ekers \& Van Drom}{1991}]{Hutsemekers} 
     Hutsem\'ekers D., Van Drom E., 1991, A\&A, 248, 141
\bibitem[\protect\citeauthoryear{Kochanek}{2011}]{Kochanek_11}      
     Kochanek C. S., 2011, ApJ, 743, 73
\bibitem[\protect\citeauthoryear{Kotak \& Vink}{2006}]{Kotak_06}
    Kotak R., Vink J. S., 2006, A\&A, 460, L5     
\bibitem[\protect\citeauthoryear{Lagage et al.}{2004}]{Lagage} 
     Lagage P.O., Pel J. W., Authier M., et al. 2004, The Messenger, 117, 12
\bibitem[\protect\citeauthoryear{Lamers et al.}{1996}]{Lamers_96}      
     Lamers H.J.G.L.M., Morris P.W., Voors R.H.M., et al., 1996, A\&A,315, 225 
\bibitem[\protect\citeauthoryear{Lamers et al.}{2001}]{Lamers_01}      
     Lamers H.J.G.L.M., Nota A., Panagia N., Smith L. J., Langer N, 2001, ApJ, 551, 764
\bibitem[\protect\citeauthoryear{Lamers \& Nugis}{2002}]{LN02}      
     Lamers H.J.G.L.M., Nugis T., 2002, A\&A, 395, L1      
\bibitem[\protect\citeauthoryear{Lau et al.}{2016}]{Lau_16}
  Lau R. M., Hankins M. J., Herter T. L., Morris M. R., Mills E. A. C. Ressler M. E, 2016, ApJ, 818, 117.     
\bibitem[\protect\citeauthoryear{Leitherer \& Robert}{1991}]{Leitherer_91}
    Leitherer C., Robert C., 1991, ApJ, 377, 6     
\bibitem[\protect\citeauthoryear{Machado et al.}{2002}]{Machado_02}
   Machado M. A. D., de Ara\'ujo F. X.., Pereira C. B., Fernandes M. B., 2002, A\&A, 387, 151
 \bibitem[\protect\citeauthoryear{Maeder \& Desjacques}{2001}]{Maeder_01}
 Maeder A., Desjacques V., 2001, A\&A, 372, L9
\bibitem[\protect\citeauthoryear{Mayer et al.}{2016}]{Mayer_16}
   Mayer A., Deschamps R., Jorissen A., 2016, A\&A, 587, 30 
\bibitem[\protect\citeauthoryear{Mihalas \& Binney}{1981}]{Mihalas}
   Mihalas D., Binney J., 1981, in Mihalas D., Binney J., eds, Galactic Astronomy: Structure and Kinematics, 2nd edn., San Francisco, CA, W. H. Freeman and Co.
\bibitem[\protect\citeauthoryear{Molinari et al.}{2010}]{Molinari_10}
   Molinari S., et al., 2010, PASP, 122, 314
\bibitem[\protect\citeauthoryear{Nota et al.}{1995}]{Nota_95}
   Nota A., Livio M., Clampin M., Schulte-Ladbeck R., 1995, ApJ, 448,788   
\bibitem[\protect\citeauthoryear{Nota et al.}{1997}]{Nota_97}
   Nota A.,Smith L., Pasquali A., Clampin M., Stroud M., 1997, ApJ, 486,338
\bibitem[\protect\citeauthoryear{Ohara et al.}{2003}]{Ohara_03}
   O'Hara T. B., Meixner M., Speck A. K., Ueta T., Bobrowsky M., 2003, ApJ, 598, 1255
\bibitem[\protect\citeauthoryear{Ottensamer et al.}{2011}]{Ottensamer_11}
   Ottensamer R., et al., 2011, in Why Galaxies Care
about AGB Stars II: Shining Examples and Common Inhabitants, eds. F.
Kerschbaum, T. Lebzelter, \& R. F. Wing, ASP Conf. Ser., 445, 625 
\bibitem[\protect\citeauthoryear{Panagia \& Felli}{1975}]{Panagia_75}
    Panagia N., Felli M., 1975, A\&A, 39, 1     
\bibitem[\protect\citeauthoryear{Poglitsch}{2010}]{Poglitsch_10}
   Poglitsch A. et al., 2010, A\&A, 518, L2
\bibitem[\protect\citeauthoryear{Smith \& Owocki}{2006}]{smow_06}
Smith N., Owocki S. P., 2006, ApJ, 645, L42
\bibitem[\protect\citeauthoryear{Smith}{2014}]{smith_14}
Smith N., 2014, ARA\&A, 52, 487 
\bibitem[\protect\citeauthoryear{Traficante et al.}{2011}]{Traficante_11} 
     Traficante A. et al., 2011, MNRAS, 416, 2932 
\bibitem[\protect\citeauthoryear{Ueta et al.}{2008}]{Ueta_08} 
     Ueta T., et al., 2008, PASJ, 60, S407
\bibitem[\protect\citeauthoryear{Umana et al.}{2005}]{Umana_05} 
     Umana G., Buemi C.S., Trigilio C., Leto P., 2005, A\&A, 437, L1     
\bibitem[\protect\citeauthoryear{Umana et al.}{2009}]{umana_09} 
     Umana G., Buemi C.S., Trigilio C., Hora J.L., Fazio G.G., Leto P., 2009, ApJ, 694, 697
\bibitem[\protect\citeauthoryear{Umana et al.}{2010}]{Umana_10} 
     Umana G., Buemi C.S., Trigilio C., Leto P., Hora J.L., 2010, ApJ, 718, 1036
\bibitem[\protect\citeauthoryear{Vamvatira-Nakou et al.}{2013}]{Vamvatira-Nakou_13} 
Vamvatira-Nakou C., Hutsem\'ekers D., Royer P., Naz\'e Y., Magain P., Exster K., Waelkens C., Groenewegen M. A. T., 2013, A\&A, 557, 20
\bibitem[\protect\citeauthoryear{Vamvatira-Nakou et al.}{2015}]{Vamvatira-Nakou_15} 
Vamvatira-Nakou C., Hutsem\'ekers D., Royer P., Cox N. L. J., Naz\'e Y., Rauw G., Waelkens C., Groenewegen M. A. T., 2015, A\&A, 578, 108
\bibitem[\protect\citeauthoryear{van Genderen et al.}{1991}]{Genderen_91} 
     van Genderen A. M., Robijn F. A. H., van Esch B. P.M., Lamers H. J. G. L. M., 1991, A\&A, 246, 4 
\bibitem[\protect\citeauthoryear{van Genderen}{2001}]{Genderen_01} 
    van Genderen A. M., 2001, A\&A, 366, 508             
\bibitem[\protect\citeauthoryear{van Leeuwen}{2007}]{Leeuwen_07}
     van Leeuwen F., 2007, A\&A, 474, 653
\bibitem[\protect\citeauthoryear{Vink \& de Koter}{2002}]{Vink_koter}    
     Vink J. S., \& de Koter A., 2002, A\&A, 393, 543
\bibitem[\protect\citeauthoryear{Voors et al.}{1997}]{Voors_97}
     Voors R.H.M., Waters L.B.F.M., Trams N.., Kaufl H.U., 1997, A\&A, 321, L21
\bibitem[\protect\citeauthoryear{Weis et al.}{1997}]{Weis_97}
     Weis K., Duschl,W. J., Bomans, D. J., Chu, Y. H., \& Joner, M. D. 1997, A\&A, 320, 568
\bibitem[\protect\citeauthoryear{Weis}{2001}]{Weis_01}
     Weis K., 2001, Rev. Modern Astron., 14, 261     
\bibitem[\protect\citeauthoryear{White}{2000}]{White_00}      
     White S. M., 2000, ApJ, 539, 851
\bibitem[\protect\citeauthoryear{Wright et al.}{2010}]{wright_10}
     Wright E. L. et al., 2010, AJ, 140, 1868
\end{thebibliography}
\end{document}